\documentclass[prb,superscriptaddress,twocolumn,floatfix,amsmath,amssymb,showpacs]{revtex4}
\usepackage{graphicx}
\usepackage{dcolumn}
\usepackage{verbatim}
\usepackage{times}
\usepackage{subfigure}
\usepackage{bm}
\usepackage{color}
\usepackage{subfigure}
\usepackage[colorlinks,bookmarks=false,citecolor=blue,linkcolor=red,urlcolor=blue,backref=red,dvipdfm]{hyperref}
\usepackage{booktabs}
\usepackage{appendix}
\begin{document}

\title{$\mathcal{Z}_2$ classification for a novel antiferromagnetic topological insulating phase in three-dimensional topological Kondo insulator}

\author{Huan Li}
\email{lihuan@glut.edu.cn}
\affiliation{College of Science, Guilin University of Technology, Guilin 541004, China}

\author{Yin Zhong}
\affiliation{Center for Interdisciplinary Studies $\&$ Key Laboratory for
Magnetism and Magnetic Materials of the MoE, Lanzhou University, Lanzhou 730000, China}

\author{Yu Liu}
\affiliation{Institute of Applied Physics and Computational Mathematics, Beijing 100088, China}
\affiliation{Software Center for High Performance Numerical Simulation, China Academy of Engineering Physics, Beijing 100088, China}

\author{Hong-Gang Luo}
\affiliation{Center for Interdisciplinary Studies $\&$ Key Laboratory for
Magnetism and Magnetic Materials of the MoE, Lanzhou University, Lanzhou 730000, China}
\affiliation{Beijing Computational Science Research Center, Beijing 100084, China}

\author{Hai-Feng Song}
\email{song\_haifeng@iapcm.ac.cn}
\affiliation{Institute of Applied Physics and Computational Mathematics, Beijing 100088, China}
\affiliation{Software Center for High Performance Numerical Simulation, China Academy of Engineering Physics, Beijing 100088, China}

\date{\today}

\begin{abstract}

Antiferromagnetic topological insulator (AFTI) is a topological matter that breaks time-reversal symmetry. Since its proposal, explorations of AFTI in strong-correlated systems are still lacking. In this paper, we show for the first time that a novel AFTI phase can be realized in three-dimensional topological Kondo insulator (TKI).
In a wide parameter region, the ground states of TKI undergo a second-order transition to antiferromagnetic insulating phases which conserve a combined symmetry of time reversal and a lattice translation, allowing us to derive a $\mathcal{Z}_2$-classification formula for these states. By calculating the $\mathcal{Z}_2$ index, the antiferromagnetic insulating states are classified into (AFTI) or non-topological antiferromagnetic insulator (nAFI) in different parameter regions. On the antiferromagnetic surfaces in AFTI, we find topologically protected gapless Dirac cones inside the bulk gap, leading to metallic Fermi rings exhibiting helical spin texture with weak spin-momentum locking.
Depending on model parameters, the magnetic transitions take place either between AFTI and strong topological insulator, or between nAFI and weak topological insulator.
By varying some model parameters, we find a topological transition between AFTI and nAFI, driving by closing of bulk gap.
Our work may account for the pressure-induced magnetism in TKI compound SmB$_6$, and helps to explore richer AFTI phases in heavy-fermion systems as well as in other strong-correlated systems.

\end{abstract}


%75.30.Mb :valence fluctuation,Kondo lattice, heavy-fermion
%75.70.Tj Spin-Orbit coupling effect
%73.20.-r   electronic surface state
%75.30.Kz   magnetic phase boundaries

\maketitle

\section{Introduction}

Kondo insulators (KI) are a series of heavy-fermion compounds, in which the hybridization between conducting $d$ electrons and local $f$ moments generates a bulk energy gap at low temperatures~\cite{Menth69}. In recent years, the proposal of topologically protected surface states in Kondo insulator SmB$_6$ has been gathering renewed interests,~\cite{Dzero10,Dzero12,Werner14,Alexandrov13,Li14,Baruselli14,Kim14,Werner13,Roy14,Chou16} and the gapless surface states inside the bulk gap have been confirmed by spin- and angle-resolved photoemission spectroscopy (SARPES),~\cite{Xu14,Xu16} which can explain the mystery plateau of resistivity at low temperatures.

Theoretically, topological Kondo insulators (TKI) can be described by the
topological periodic Anderson model (T-PAM) with a spin-orbit coupled nonlocal hybridization between $d$ and $f$ orbits.~\cite{Dzero10,Dzero12} The spin-orbit-coupled nature and odd parity of such hybridization guarantee both time-reversal symmetry (TRS) and space-inversion symmetry of T-PAM, leading to a $\mathcal{Z}_2$ topological classification of the insulating phases.~\cite{Fu07,Dzero10}  According to the relative altitude between $d$ and renormalized $f$ levels at eight time-reversal invariant momenta (TRIM) in 3D Brillouin zone (BZ), insulating states can be classified into strong topological insulator (STI), weak topological insulator (WTI) and non-topological Kondo insulator (nKI),~\cite{Dzero10} in which STI and WTI exhibit gapless surface Dirac cones driven by band inversions between $d$ and $f$ bands at certain TRIM.~\cite{Dzero12}
Since the effective $f$ level is strongly renormalized from the bare one $\epsilon_f$, the variations of hybridization strength $V$ and $\epsilon_f$ can cause successive topological transitions among STI, WTI and nKI,~\cite{Tran12} moreover, with different choices of electron-hopping amplitudes (EHA), the topological transition processes are quite distinct.~\cite{Legner14}

In a Kondo lattice at or near half-filling, antiferromagnetic (AF) order is quite favored in relative weak Kondo interaction region,~\cite{Si14,Li15}
while in ordinary half-filled periodic Anderson model (PAM) with local $d$-$f$ hybridization, it is well-known that a ground-state magnetic transition occurs from AF to paramagnetic (PM) phase by enhanced hybridization, as revealed by numerical simulations and theoretical calculations.~\cite{Vekic95,Leo08,Horiuchi08,Watanabe09,Yang93,Sun93,Sun95}
For a TKI compound SmB$_6$, literature has witnessed an emergence of magnetic order under high pressure.~\cite{Barla05,Derr06,Derr08,Nishiyama13,Paraskevas15,Butch16,Emi16,Zhao17}
Similar to normal Kondo lattice systems, the pressure-induced magnetism in SmB$_6$ may be naturally considered to be AF ordered,~\cite{Chang17} although no clear experimental evidence is present to date. The topological essence of SmB$_6$ naturally reminds us of the possible non-trivial topology of the induced AF phase. Indeed, in some low-dimensional topological Kondo lattices, AF phases with non-trivial topologies have been verified theoretically.~\cite{Zhong13} These considerations motivate us to study
the transition to AF phases in three-dimensional (3D) TKI, to classify these AF states and search topologically protected AF phase, then further reveal the transition processes among AF states and various PM topological insulating (TI) phases.

In a TI, once AF order sets in, TRS is broken, then the standard $\mathcal{Z}_2$ classification is no longer applicable to AF phases. Nevertheless, Mong et al. have pointed out that if there is a translation $\mathcal{T}_{\mathbf{D}}$ by lattice vector $\mathbf{D}$ which inverts the AF magnetization at all sites, then such AF state is invariant under a combined operation $\mathcal{S}=\Theta\mathcal{T}_{\mathbf{D}}$, since time-reversal operation $\Theta$ also inverts the magnetization.~\cite{Mong10,Fang13,Zhang15}
$\mathcal{S}$ is antiunitary, and $\mathcal{S}^2$= -1 at four out of the eight high-symmetric points in 3D magnetic Brillouin zone (MBZ), ensuring Kramers degeneracy at these four momenta (KDM).~\cite{Fang13} These properties lead to a new type of $\mathcal{Z}_2$ classification of these AF insulating states, in which an antiferromagnetic topological insulating phase (AFTI) arises by adding weak staggered magnetization into STI phase while maintaining an insulating bulk gap.~\cite{Mong10} Since the proposal, a few AFTI phases have been suggested in some non-interacting models in which the AF orders are added artificially by Zeeman terms,~\cite{Essin12,Baireuther14,Begue17,Ghosh17} or in low-dimensional models with electron correlation.~\cite{Miyakoshi13} For a 3D AFTI candidate GdBiPt,~\cite{Muller14} an AF phase has been suggested,~\cite{ZhiLi15} but it is not a full-gapped insulating state. The explorations of AFTI in real systems, particularly in 3D strong-correlated systems in which AF orders can arise naturally, are still lacking.

The magnetism observed in pressured SmB$_6$~\cite{Barla05,Derr06,Derr08,Nishiyama13,Paraskevas15,Butch16,Emi16,Zhao17} sheds light on possible emergence of AFTI in 3D TKI. Recently, Peters et al. reported their study on bulk and surface magnetism in TKI by dynamic mean-field theory (DMFT), but the topologies of the magnetic states were not deduced explicitly;~\cite{Peters18} Chang et al. proposed an interesting topological A-AF phase which is staggered along single axis in pressured SmB$_6$ by first-principle calculations,\cite{Chang17} in which the strong correlation was not considered explicitly. Moreover, the AF phases suggested by Chang et al. are actually bulk-conducting states, essentially different from the topologically protected insulating AFTI state which requires a full bulk gap. Therefore, the possible AFTI in 3D TKI still lacks investigation.

As mentioned above, with different model parameters, distinct TI phases emerge in TKI in weak hybridization region,~\cite{Legner14} in which magnetic transitions may take place to induce AF states. Consequently, once TKI undergoes magnetic transitions, the induced AF orders grow from distinct TI phases, depending on the parameter regions, leading to topologically distinguishable AF phases, i.e., an AF order growing from STI leads to an AFTI state,~\cite{Mong10} if AF order grows from WTI or nKI, a non-topological AF insulator (nAFI) arises.
Furthermore, if the magnetic transition occurs near the phase boundary between STI and WTI, varying the model parameters in a special way may induce a topological transition between AFTI and nAFI.
Therefore, in TKI, the existence of AFTI and nAFI, the topological-classification formula for them, and possible topological transition between AFTI and nAFI highly deserve investigations in a self-consistently manner from the original T-PAM model for TKI.

In this work, we verify the existence of AFTI in 3D TKI for the first time. We explore the transition to insulating AF states in 3D TKI, and present a $\mathcal{Z}_2$ topological formula to classification these AF states. In some parameter region, we find a novel AFTI with topologically-protected gapless surface Dirac cones, exhibiting helical spin texture on the Fermi rings. We also obtain a topological transition from AFTI to nAFI. To our knowledge, the novel AFTI, and the topological transition between AFTI and nAFI in 3D TKI are reported for the first time by our work.

This paper is arranged as follows. In section II, we adopt a mean-field Kotliar-Ruckenstein (K-R) slave-boson representation~\cite{Kotliar86,Yang93,Sun93,Sun95} for T-PAM to describe the possible AF phases in 3D TKI in general parameter region. The AF configuration studied in this paper is staggered by adjacent sites, and meets the requirement of $\mathcal{S}$-invariance.
In section III, we study the topological transitions among STI, WTI and nKI, and select two typical transition processes by choosing two sets of EHA, which are appropriate for studying AF transitions and will lead to topologically distinguishable AF phases.
In section IV, by locating the four KDM, we derive the expression for $\mathcal{Z}_2$ index in AF states, which is determined by the product of parities at these KDM.
In section V, we perform a saddle-point solution for AF phase to derive the evolution of staggered magnetization and determine the magnetic transition points on $\epsilon_f$-$V$ plane. By calculating the $\mathcal{Z}_2$ index, we find that the two sets of EHA in section III lead to two topologically distinguishable AF phases, one is AFTI evolving from STI, and the other is nAFI evolving from WTI.
By diagonalizing parallel slabs with (001) surface, we observe the expected gapless surface Dirac cones with helical spin texture in AFTI, confirming its non-trivial topology.
In section VI, we demonstrate a topological transition between AFTI and nAFI driving by closing and reopening of bulk gap, and an insulator-metal transition inside the AF phases, then summarize the phase transitions in a global phase diagram.

Our work provides deeper understanding of AFTI in 3D TKI, and may account for the magnetism observed in pressured SmB$_6$. Our algebra also helps to investigate richer AFTI phases in heavy-fermion systems, as well as in other strong-correlated systems.

\section{K-R representation for AF states}

We consider a simplified model for 3D TKI: half-filled spin-1/2 T-PAM with spin-orbit coupled hybridization between neighboring $d$ and $f$ electrons in cubic lattice, with lattice constant $a$ and $N$ sites:~\cite{Alexandrov15}
\begin{align}
\mathcal{H}=&\sum_{i,j,\sigma}(t^d_{ij}d^\dag_{i\sigma}d_{j\sigma}+t^f_{ij}f^\dag_{i\sigma}f_{j\sigma})
+\epsilon_f\sum_{i,\sigma}f^\dag_{i\sigma}f_{i\sigma}\nonumber\\
+&U\sum_in^f_{i\uparrow}n^f_{i\downarrow}
-(\frac{\textrm{i}V}{2}\sum_{i,\vec{l},\alpha,\beta}\vec{l}\cdot\vec{\sigma}_{\alpha\beta}d^\dag_{i\alpha}f_{i+\vec{l},\beta}+h.c.)\nonumber\\
-&\mu\sum_{i,\sigma}(d^\dag_{i\sigma}d_{i\sigma}+f^\dag_{i\sigma}f_{i\sigma}),
\label{PAM}\end{align}
where $d^\dag_{i\sigma}$ and $f^\dag_{i\sigma}$ create a $d$ or $f$ electron at site $i$, with spin $\sigma=\pm1$ representing spin up and down, respectively. The electron hopping amplitudes (EHA) include nearest-neighbor (NN) hopping, next-nearest-neighbor (NNN) hopping, and next-next-nearest-neighbor (NNNN) hopping for both $d$ and $f$ electrons, and we set $t^d_{ij}=-t_d$, $t^f_{ij}=-t_f$ for NN, $t^d_{ij}=-t^\prime_d$, $t^f_{ij}=-t^\prime_f$ for NNN, and $t^d_{ij}=-t^{\prime\prime}_d$, $t^f_{ij}=-t^{\prime\prime}_f$ for NNNN. In what follows, $t_d=1$ is chosen as energy unit, and we consider $t_f/t_d<0$ and keep $t^\prime_d /t_d$ close to $t^\prime_f /t_f$, and $t^{\prime\prime}_d /t^\prime_d$ close to $t^{\prime\prime}_f /t^\prime_f$, to get a full bulk gap,~\cite{Legner14,Alexandrov15,Zhong17} since we focus on the insulating phases at half-filling.
$\mu $ denotes the chemical potential, which fits the total electron number per site to $n_t=2$. $\epsilon_f$ is the energy level of $f$ orbit, $U$ is the on-site Coulomb repulsion between $f$ electrons. $\vec{l}'s$ are the six coordination vectors in cubic lattice.
$\vec{\sigma}$ stands for the Pauli matrix.
In TKI, the opposite parities of $d$ and $f$ orbits cause the hybridization to acquire a non-local spin-dependent form, which is explicitly written as~\cite{Alexandrov15}
\begin{align}
\mathcal{H}_{df}=&-\frac{\textrm{i}V}{2}\sum_{i,\pm,\sigma}(\pm d^\dag_{i\sigma}f_{i\pm {\mathbf{a}_1},\bar{\sigma}}\mp \textrm{i}\sigma d^\dag_{i\sigma}f_{i\pm {\mathbf{a}_2},\bar{\sigma}}\pm \sigma d^\dag_{i\sigma}f_{i\pm {\mathbf{a}_3},\sigma})\nonumber\\
&+h.c.,
\end{align}
with $\bar{\sigma}=-\sigma$ and $\mathbf{a}_1$, $\mathbf{a}_2$, $\mathbf{a}_3$ are three lattice basic vectors along $x$, $y$, $z$ axis, respectively.
$\mathcal{H}_{df}$ can be written in momentum space by~\cite{Alexandrov15}
\begin{align}
\mathcal{H}_{df}=V\sum_{\mathbf{k},\alpha,\beta}\mathbf{S}_{\mathbf{k}}\cdot \vec{\sigma}_{\alpha\beta}d^\dag_{\mathbf{k}\alpha}f_{\mathbf{k}\beta}+h.c.,
\end{align}
where the vector $\mathbf{S}_\mathbf{k}$=$(\sin k_x,\sin k_y,\sin k_z)$,with $k_x=\mathbf{k}\cdot\mathbf{a}_1$, $k_y=\mathbf{k}\cdot\mathbf{a}_2$, $k_z=\mathbf{k}\cdot\mathbf{a}_3$.
Note that the particle-hole symmetry in T-PAM (Eq. (\ref{PAM})) has been broken even at half-filling, so the chemical potential term should be included in any case. In this paper, we focus on the large-$U$ limit and consider variable $\epsilon_f$, $V$ and EHA, which are appropriate to describe TKI compound SmB$_6$ with mixed valence.

It's well known that a half-filled PAM is usually AF-ordered in weak hybridization region,~\cite{Vekic95,Leo08,Horiuchi08,Watanabe09,Yang93,Sun93,Sun95} besides, in some topological Kondo lattices, AF orders are also favored.~\cite{Zhong13,Chang17}
For pressured SmB$_6$, an interesting topological A-AF state which is staggered along single axis has been suggested by first-principle calculations,~\cite{Chang17} unfortunately, such A-AF configuration is found to be unstable within our method, which may be ascribed to the large-$U$ limit we apply, while the Coulomb correlation was not considered explicitly in Ref. \onlinecite{Chang17}.
Under this context, we consider the more common AF state which is staggered between adjacent sites, and other AF configurations including A-AF state will not be considered explicitly, but we will still provide a brief topological description for these states.

The slave-boson technique has been widely used to study the TI phases and topological transitions in TKI,~\cite{Tran12,Alexandrov13,Legner14,Alexandrov15,Chou16} besides, Kotliar-Ruckenstein (K-R) slave-boson method has been applied successfully to include magnetic orders in studying the Hubbard model and PAM.~\cite{Kotliar86,Yang93,Sun93,Sun95}
In order to treat both PM and AF phases in TKI, we adopt the K-R slave-boson technique. Firstly the Hilbert space of $f$-electrons is decomposed into single occupancy, double occupancy and empty state, with two relations $\sum_\sigma P^\dag_{i\sigma}P_{i\sigma}+D^\dag_i D_i+e^\dag_i e_i=1$ and $f^\dag_{i\sigma}f_{i\sigma}=P^\dag_{i\sigma}P_{i\sigma}+D^\dag_i D_i$, which are imposed by two Lagrange terms with Lagrange multipliers $\lambda^{(1)}_i$ and $\lambda^{(2)}_{i\sigma}$, respectively. In order to reproduce correct $f$-electron occupancy, each $f$-operator $f_{\sigma}$ ($f^\dag_{\sigma}$) in $d$-$f$ hybridization and $f$-$f$ hopping terms is multiplied by a renormalization factor $Z_{i\sigma}$ ($Z^*_{i\sigma}$). By the definitions $P^2_{i\uparrow}+P^2_{i\downarrow}+2D^2_i=n^f_i$, $P^2_{i\uparrow}-P^2_{i\downarrow}=m^f_i$, $(\lambda^{(2)}_{i\uparrow}+\lambda^{(2)}_{i\downarrow})/2=\eta_i$ and
$(\lambda^{(2)}_{i\uparrow}-\lambda^{(2)}_{i\downarrow})/2=-h_i$, where $n^f_i$ is $f$-electron number, $m^f_i$ and $h_i$ are staggered order parameters at site $i$, and $\eta_i$ renormalizes the $f$ level,
the Hamiltonian Eq. (\ref{PAM}) is then rewritten as
\begin{align}\mathcal{H}=&\sum_i(h_im^f_i-\eta_in^f_i)+\sum_{\mathbf{k},\sigma}(\epsilon_\mathbf{k}-\mu)d^\dag_{\mathbf{k}\sigma}d_{\mathbf{k}\sigma}\nonumber\\
+&\sum_{i,\sigma}(\epsilon_f+\eta_i-\sigma h_i-\mu)f^\dag_{i\sigma}f_{i\sigma}+\sum_{i,j,\sigma}t^f_{ij}Z^*_{i\sigma}Z_{j\sigma}f^\dag_{i\sigma}f_{j\sigma}\nonumber\\
-&[\frac{\textrm{i}V}{2}\sum_{i,\sigma,\pm}(\pm d^\dag_{i\sigma}f_{i\pm {\mathbf{a}_1},\bar{\sigma}}Z_{i\pm {\mathbf{a}_1},\bar{\sigma}}\mp \textrm{i}\sigma d^\dag_{i\sigma}f_{i\pm {\mathbf{a}_2},\bar{\sigma}}Z_{i\pm {\mathbf{a}_2},\bar{\sigma}}\nonumber\\
&\pm \sigma d^\dag_{i\sigma}f_{i\pm {\mathbf{a}_3},\sigma}Z_{i\pm {\mathbf{a}_3},\sigma})+h.c.],\label{K-R}
\end{align}
in which we have set $U\rightarrow\infty$ then double occupance is excluded, and $\lambda^{(1)}_i$ term vanishes through mean-field approximation. For our considered AF phases, we decompose the cubic lattice into two sublattices A and B (both are face-centered cubic lattices), and by using the mean-field approximation $n^f_i=n_f$, $m^f_i=(-1)^im_f$, $h_i=(-1)^ih$, we have $Z_{A\uparrow}=Z_{B\downarrow}=Z_1$, and
$Z_{A\downarrow}=Z_{B\uparrow}=Z_2$, with
\begin{align}
Z_1=\sqrt{\frac{2(1-n_f)}{2-n_f-m_f}},
Z_1=\sqrt{\frac{2(1-n_f)}{2-n_f+m_f}}.
\end{align}

Through such mean-field treatment, we obtain the effective Hamiltonian in momentum space with a matrix form
\begin{equation}
\mathcal{H}=N(hm_f-\eta n_f)+\sum_{\mathbf{k}\in \mathrm{MBZ}}\Psi _{
\mathbf{k} }^{\dag }\mathbf{H}_{\mathbf{k} }\Psi _{\mathbf{k}
 },
\label{Hamiltonian}\end{equation}
where the summation of $\mathbf{k}$ is restricted in
the magnetic Brillouin zone (MBZ). A eight-component operator is defined as $\Psi _{\mathbf{k}}=(d_{\mathbf{k}A\uparrow},d_{\mathbf{k}A\downarrow},
d_{\mathbf{k}B\uparrow},d_{\mathbf{k}B\downarrow},
f_{\mathbf{k}A\uparrow},f_{\mathbf{k}A\downarrow},
f_{\mathbf{k}B\uparrow},f_{\mathbf{k}B\downarrow})^T$, and the Hamiltonian matrix is given by
 \begin{align}
 \mathbf{H}_{\mathbf{k}}=\left(
\begin{array}{cc}
\mathbf{H}^d_{\mathbf{k}} & \mathbf{V}_{\mathbf{k}} \\
\mathbf{V}^+_{\mathbf{k}} & \mathbf{H}^f_{\mathbf{k}}
\end{array}
\right) \label{Hk},
 \end{align}
 with
 \begin{align}
 \mathbf{H}^d_{\mathbf{k}}=\left(
\begin{array}{cc}
(t^\prime_d\gamma_{\mathbf{k}}-\mu)I_{2} &  u^d_{\mathbf{k}}I_{2}\\
  u^d_{\mathbf{k}}I_{2} & (t^\prime_d\gamma_{\mathbf{k}}-\mu)I_{2}
\end{array}
\right),
 \end{align}
 \begin{align}
 \mathbf{V}_{\mathbf{k}}=V\left(
\begin{array}{cccc}
0&0&Z_2\sin k_z&Z_1\Gamma_{\mathbf{k}}\\
0&0&Z_2\Gamma^*_{\mathbf{k}}&-Z_1\sin k_z\\
Z_1\sin k_z&Z_2\Gamma_{\mathbf{k}}&0&0\\
Z_1\Gamma^*_{\mathbf{k}}&-Z_2\sin k_z&0&0
\end{array}
\right),
 \end{align}
 \begin{align}
& \mathbf{H}^f_{\mathbf{k}}=
 &\left(
\begin{array}{cccc}
e_{1\mathbf{k}}&0&Z_1Z_2u^f_{\mathbf{k}}&0\\
0&e_{2\mathbf{k}}&0&Z_1Z_2u^f_{\mathbf{k}}\\
Z_1Z_2u^f_{\mathbf{k}}&0&e_{2\mathbf{k}}&0\\
0&Z_1Z_2u^f_{\mathbf{k}}&0&e_{1\mathbf{k}}
\end{array}
\right),
\end{align}
where $I_2$ is a two-order unit matrix, $\Gamma_{\mathbf{k}}=\sin k_x-\textrm{i}\sin k_y$, $\gamma_{\mathbf{k}}=-4(\cos {k_{x}}\cos {k_{y}}+\cos {k_{x}}\cos {k_{z}}+\cos {k_{y}}\cos {k_{z}})$,  $e_{1\mathbf{k}}=\epsilon_f+\eta-\mu-h+t^\prime_fZ^2_1\gamma_{\mathbf{k}}$, $e_{2\mathbf{k}}=\epsilon_f+\eta-\mu+h+t^\prime_fZ^2_2\gamma_{\mathbf{k}}$,
$u^{d(f)}_{\mathbf{k}}=t_{d(f)}\lambda_{\mathbf{k}}+t^{\prime\prime}_{d(f)}g_{\mathbf{k}}$ with $\lambda _{\mathbf{k}}=-2(\cos {k_{x}}+\cos {k_{y}}+\cos {k_{z}})$ and $g_{\mathbf{k}}=-8\cos {k_{x}}\cos {k_{y}}\cos {k_{z}}$.
In general case, the Hamiltonian matrix Eq. (\ref{Hk}) should be diagonalized numerically to obtain the four branches of dispersions $E^{(i)}_\mathbf{k}$ ($i=1,2,3,4$) which are all two-fold degenerate.
The mean-field parameters $n_f$, $m_f$, $h$, $\eta$ and the chemical potential $\mu$ in AF phase should be determined by saddle-point equations self-consistently, which are given in appendix A, and the numerical solution of these equations will be performed later in section V.
With the calculated parameters, we can compute the ground-state energy of AF phase by
\begin{align}
E^{AF}_g=N(hm_f-\eta n_f+\mu n_t)
+\sum_{\mathbf{k}}\sum_{i=1,..,4}\theta(-E^{(i)}_\mathbf{k})E^{(i)}_\mathbf{k}.\label{FAF}
\end{align}

\section{Topological transitions between topological insulating phases}

Before studying the magnetic transitions in TKI, we consider the topological transitions among various TI phases, then draw the topological phase diagrams, in order to select appropriate model parameters to further discuss magnetic transitions and AF phases. The K-R slave boson scheme for PM phases follows from Eq. (\ref{K-R}), in which $m_f$ and $h$ should be eliminated, to arrive at the mean-field Hamiltonian in large-$U$ limit
\begin{equation}
\mathcal{H}=N(-\eta n_f)+{\sum_{\mathbf{k}}}\Psi _{
\mathbf{k} }^{\dag }\mathbf{H}_{\mathbf{k} }\Psi _{\mathbf{k}
 },
\label{PM}\end{equation}with
$\Psi _{\mathbf{k}}=(d_{\mathbf{k}\uparrow}d_{\mathbf{k}\downarrow}
f_{\mathbf{k}\uparrow}f_{\mathbf{k}\downarrow}
)^T$ and
\begin{align}
 \mathbf{H}_{\mathbf{k}}=\left(
\begin{array}{cc}
(\epsilon^d_{\mathbf{k}}-\mu)I_2&\mathbf{V}_{\mathbf{k}}\\
\mathbf{V}^+_{\mathbf{k}}&(\tilde{\epsilon}^f_{\mathbf{k}}-\mu)I_2
\end{array}\label{VZ}
\right),\end{align}
in which the renormalized $f$ dispersion $\tilde{\epsilon}^f_{\mathbf{k}}=\epsilon_f+\eta+Z^2\epsilon^f_{\mathbf{k}}$, the renormalization factor $Z=\sqrt{2(1-n_f)/(2-n_f)}$, the effective hybridization
\begin{align}
 \mathbf{V}_{\mathbf{k}}=VZ\left(
\begin{array}{cc}
\sin k_z&\sin k_x-\textrm{i}\sin k_y\\
\sin k_x+\textrm{i}\sin k_y&-\sin k_z
\end{array}
\right),
 \end{align}and tight-binding dispersions $\epsilon^{d(f)}_{\mathbf{k}}=t_{d(f)}\lambda_{\mathbf{k}}+t^\prime_{d(f)}\gamma_{\mathbf{k}}+t^{\prime\prime}_{d(f)}g_{\mathbf{k}}$.
The ground-state energy is then $E^{PM}_g=N(-\eta n_f+\mu n_t)+2\sum_{\mathbf{k},\pm}\theta(-E^\pm_\mathbf{k})E^\pm_\mathbf{k}$, with quasi-particle dispersions
\begin{align}
E^\pm_\mathbf{k}=\frac{1}{2}[\epsilon^d_\mathbf{k}+\tilde{\epsilon}^f_\mathbf{k}
\pm\sqrt{(\epsilon^d_\mathbf{k}-\tilde{\epsilon}^f_\mathbf{k})^2+4V^2Z^2S^2_\mathbf{k}}]-\mu,
\label{Eplusminus}\end{align}
which are both two-fold degenerate. $n_f,\mu$ and $\eta$ are computed through saddle-point solution of $E^{PM}_g$, see appendix B.

\heavyrulewidth=1bp

\begin{table}
\small
\renewcommand\arraystretch{1.3}
%\centering
\caption{\label{EHA}
Two sets of EHA used in this work$^a$}
%\begin{ruledtabular}
\begin{tabular*}{8cm}{@{\extracolsep{\fill}}ccccccc}
\toprule
     &$t_d$ & $t^\prime_d$ & $t^{\prime\prime}_{d}$  & $t_f$ & $t^\prime_f$ & $t^{\prime\prime}_{f}$ \\
\hline
EHA(I)&1 & 0.15 & 0 & -0.2 & -0.02 & 0 \\
EHA(II)&1 & -0.375 & -0.375 & -0.2 & 0.09 & 0.09 \\
\bottomrule
\end{tabular*}
%\end{ruledtabular}
\footnotetext[1]{In section VI, we also use other EHA.}
\end{table}

\begin{figure}[tbp]
\hspace{-0.1cm} \includegraphics[totalheight=2.63in]{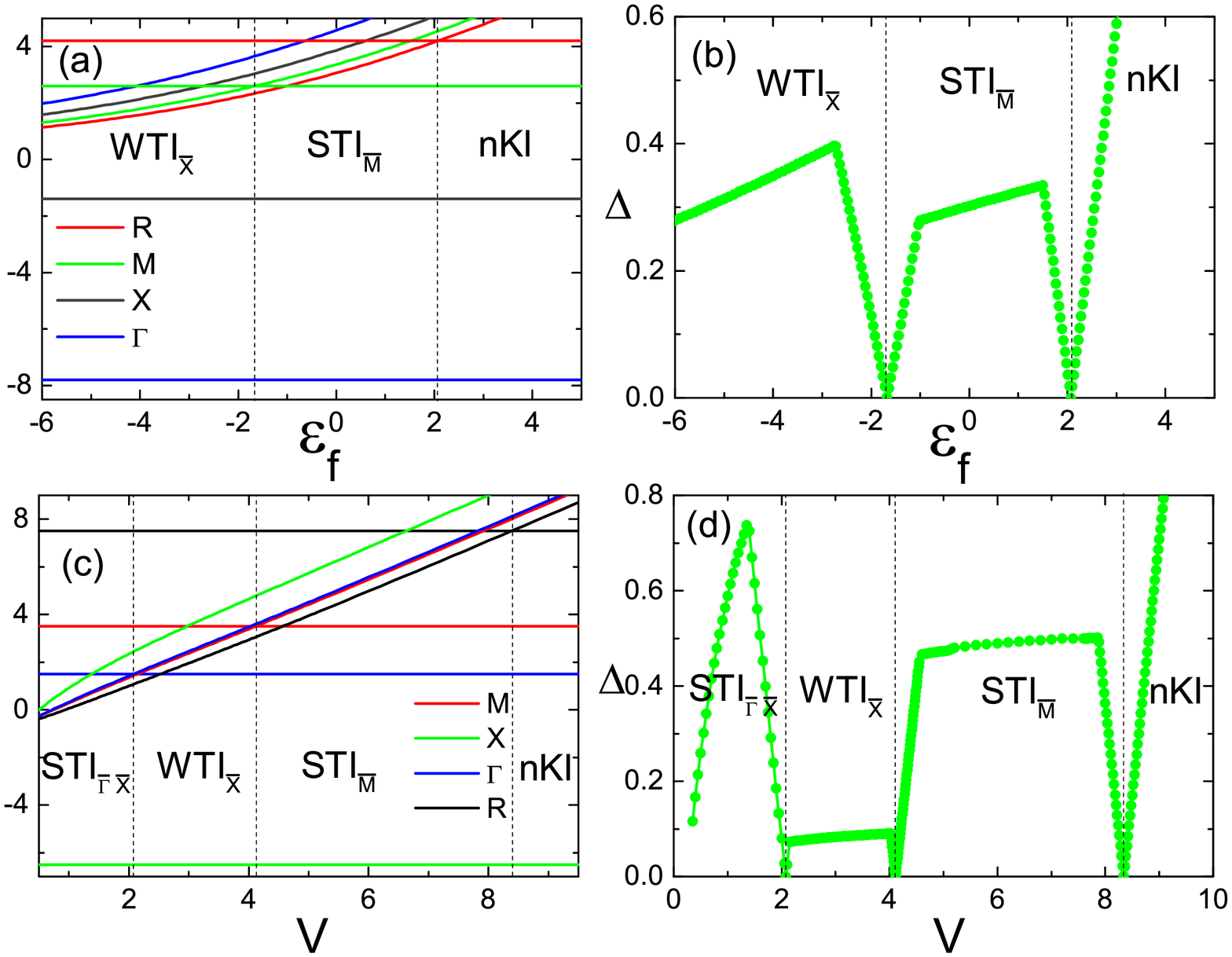}
\caption{(Color online)
$d$ dispersion $\epsilon^d_{\mathbf{k}_m}$ and renormalized $f$ dispersion $\tilde{\epsilon}^f_{\mathbf{k}_m}$ at TRIM $\mathbf{k}_m$ vs (a) bare $f$-level $\epsilon_f$ or (c) hybridization strength $V$. Model parameters are $V=3$ with EHA(I) for (a), and $\epsilon_f=-2$ with EHA(II) for (c), respectively. In (a) and (c), the flat lines denote $\epsilon^d_{\mathbf{k}_m}$, while the curves show $\tilde{\epsilon}^f_{\mathbf{k}_m}$. The classification of TI phases is governed by $Z_2$ invariants, see the text.
(b) and (d): Corresponding bulk gap evolutions in (a) and (c), respectively, showing gap-closing at topological transitions.
}
\label{fig1}
\end{figure}
The spin-orbit-coupled nature of effective hybridization $\mathbf{V}_{\mathbf{k}}$ guarantees TRS of Hamiltonian Eq. (\ref{PM}): $[\Theta,\mathcal{H}]=0$ or equivalently
$\Theta\mathbf{H}_{\mathbf{k}}\Theta^{-1}$=$\mathbf{H}_{-\mathbf{k}}$, and the existence of inversion center ensures an additional space-inversion symmetry: $[\mathcal{P},\mathcal{H}]=0$ or $\mathcal{P}\mathbf{H}_{\mathbf{k}}\mathcal{P}^{-1}=\mathbf{H}_{-\mathbf{k}}$, in which time-reversal operator $\Theta=\mathrm{i}(I_2\otimes\sigma_y)\mathcal{K}$ ($\mathcal{K}$ is complex conjugation) and the parity matrix $\mathcal{P}=\sigma_z\otimes I_2$ in present basis.
$\Theta$ is anti-unitary with $\Theta^2=-1$, leading to Kramers degeneracy at eight TRIM in 3D BZ, resulting in $\mathcal{Z}_2$ classification of the insulating states.
The $\mathcal{Z}_2$ invariants are calculated by parities of the occupied states at eight TRIM $\mathbf{k}_m$ in the BZ.~\cite{Fu07,Dzero10} We use standard notation to denote $\mathbf{k}_m$: $\Gamma$ for $(0,0,0)$,  $X$ for $(\pi,0,0)$, $(0,\pi,0)$, $(0,0,\pi)$, $M$ for $(\pi,\pi,0)$, $(\pi,0,\pi)$, $(0,\pi,\pi)$, and $R$ for $(\pi,\pi,\pi)$. For the 2D BZ of (001) surface, on which the surface states will be considered, there are four TRIM $\mathbf{p}_m$ denoted by $\bar{\Gamma}=(0,0)$, $\bar{X}=(\pi,0),(0,\pi)$, and $\bar{M}=(\pi,\pi)$.
Due to its odd-parity, the hybridization vanishes at TRIM, therefore at $\mathbf{k}_m$, the occupied dispersion $E^-_{\mathbf{k}_m}$=$\mathrm{min}(\epsilon^d_{\mathbf{k}_m},\tilde{\epsilon}^f_{\mathbf{k}_m})$-$\mu$ (see Eq. (\ref{Eplusminus})). Since $d$ ($f$) orbit has even (odd) parity, the parity at $\mathbf{k}_m$ is expressed by (each pair of Kramers-degenerate states at $\mathbf{k}_m$ is calculated once) $\delta_m=-\mathrm{sgn}(\epsilon^d_{\mathbf{k}_m}-\tilde{\epsilon}^f_{\mathbf{k}_m})$,~\cite{Fu07,Tran12}
then the strong topological index $\nu_0$ is related to the product of all eight $\delta_m$ by $(-1)^{\nu_0}=\prod_m\delta_m$, while the weak topological index $\nu_j$ ($j$=1,2,3) is related to the product of four $\delta_m$ on corresponding high-symmetry plane: $(-1)^{\nu_j}=\prod_{\mathbf{k}_m\in P_j}\delta_m$, in which $P_1$, $P_2$, $P_3$ denote $k_x$=0, $k_y$=0, $k_z$=0 planes, respectively. $\nu_0=1$ corresponds to a STI with odd number of surface Dirac cones,
while $\nu_0=0$ and $\nu_j=1$ indicate a WTI with even number of surface Dirac cones.~\cite{Dzero12}

\begin{figure}[tbp]
\hspace{-0.1cm} \includegraphics[totalheight=2.42in]{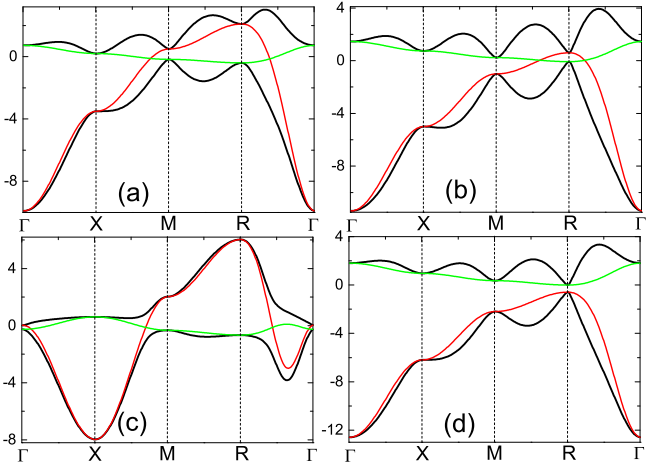}
\caption{(Color online) Bulk spectrums (black lines) vs $d$ dispersions (red lines) and renormalized $f$ dispersions (green lines) in four typical phases: (a) WTI$_{\bar{X}}$, (b) STI$_{\bar{M}}$, (c) STI$_{\bar{\Gamma}\bar{X}}$ and (d) nKI. Model parameters: (a) $V=2.8$, $\epsilon_f=-3$; (b) $V=3.5$, $\epsilon_f=-0.1$; (c) $V=1.8$, $\epsilon_f=-2$; (d) $V=3$, $\epsilon_f=3$; EHA(I) for (a), (b) and (d); EHA(II) for (c). Band-inversions are clear in WTI$_{\bar{X}}$, STI$_{\bar{M}}$ and STI$_{\bar{\Gamma}\bar{X}}$.
}
\label{fig2}
\end{figure}

\begin{figure}[tbp]
\hspace{-0cm} \includegraphics[totalheight=3.5in]{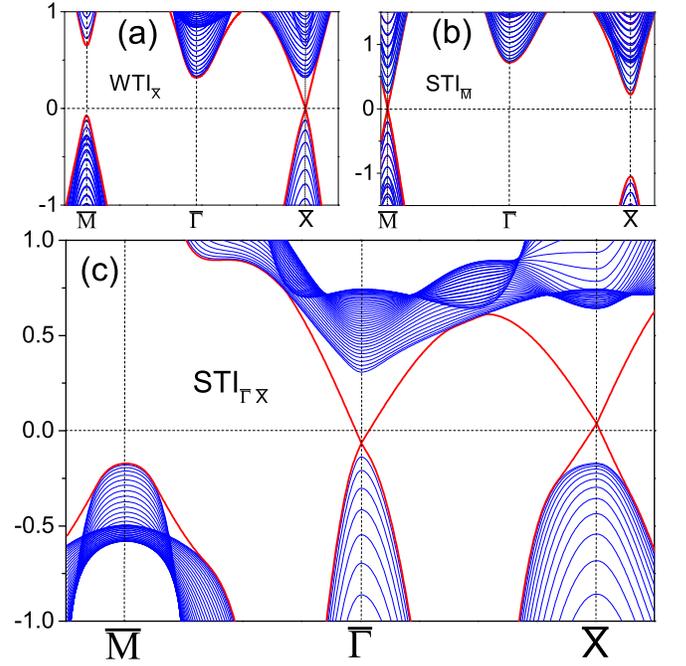}
\caption{(Color online)
Bulk spectrums (blue lines) and surface dispersions (red lines) of 40 slabs with (001) surface for (a) WTI$_{\bar{X}}$, (b) STI$_{\bar{M}}$ and (c) STI$_{\bar{\Gamma}\bar{X}}$. Except for $V=1.708$ in (c), other parameters are set according to those in Fig. \ref{fig2}. Note the small Fermi surfaces around $\bar{\Gamma}$ and $\bar{X}$ in STI$_{\bar{\Gamma}\bar{X}}$, while in WTI$_{\bar{X}}$ and STI$_{\bar{M}}$, the Dirac points cross the Fermi level.
}
\label{fig3}
\end{figure}

We choose two sets of EHA showing in Tab. \ref{EHA} to depict the topological transitions as functions of bare $f$ level $\epsilon_f$ and hybridization strength $V$, and these two sets of EHA are denoted by EHA(I) and EHA(II) in the following, respectively. The reason to choose positive (negative) $t^\prime_d$ in EHA(I) (EHA(II)) is that with physical $\epsilon_f$ and $V$ (e.g., $\epsilon_f$=-2 and $V$=1), a small positive $t^\prime_d$ leads to a WTI phase, while $t^\prime_d<-0.3$ usually leads to a STI phase. Moreover, in EHA(I), for simplicity, we choose $t^{\prime\prime}_d=t^{\prime\prime}_f=0$, since small $t^{\prime\prime}_d$ and $t^{\prime\prime}_f$ do not shift the topology. Actually, if $t^\prime_d$ varies continuously (e.g., from $-t_d$ to $t_d$), all possible TI phases in TKI can be induced on $V$-$t^\prime_d$ or $\epsilon_f$-$t^\prime_d$ planes.~\cite{Legner14}
Since our work focuses on the AF phases in TKI, we select proper hopping amplitudes, and the PM phases will not be investigated deeply.
We will show later that EHA(I) and EHA(II) will induce two topologically-distinct AF phases.

In Fig. \ref{fig1}(a), we display our numerical solutions of $d$ dispersion $\epsilon^d_{\mathbf{k}}$ and renormalized $f$ dispersion $\tilde{\epsilon}^f_{\mathbf{k}}$ at eight TRIM, as functions of $\epsilon_f$ at fixed $V=3$ with EHA(I), and corresponding bulk gap evolution is shown in Fig. \ref{fig1}(b).
For EHA(II), the results are displayed in Figs. \ref{fig1}(c) and (d), now as functions of $V$ at fixed $\epsilon_f=-2$.
As $\epsilon_f$ or $V$ varies, once $\delta_m$ changes sign at a certain $\mathbf{k}_m$, a topological phase transition occurs, denoting by the dotted vertical lines in Fig. \ref{fig1}, and the topologies on both sides should be determined by calculating $\nu_0$ and $\nu_j$.
For EHA(I), two topological transitions occur with descending $\epsilon_f$: nKI-STI transition at $\epsilon_f=2.07$ and STI-WTI transition at $\epsilon_f=-1.68$.
Due to large positive value of $\eta$ at deep $\epsilon_f$, $\tilde{\epsilon}^f_{\mathbf{k}}$ is highly renormalized, showing a saturated tendency, therefore no additional transition takes place after STI-WTI transition when $\epsilon_f$ further descends.
For EHA(II), in addition to nKI-STI and STI-WTI transitions, there is a WTI-STI transition when further reducing $V$.
At these topological transition points, since $\epsilon^d_{\mathbf{k}_m}$ equals $\tilde{\epsilon}^f_{\mathbf{k}_m}$ at certain $\mathbf{k}_m$, at which the hybridization $\mathbf{V}_{\mathbf{k}_m}$ also vanishes, one can check from Eq. (\ref{Eplusminus}) that $E^+_{\mathbf{k}_m}=E^-_{\mathbf{k}_m}$, namely the bulk gap is closed linearly towards these transition points, then is reopened thereafter, therefore the topological transitions are driven by bulk gap-closing, otherwise the topologies will be protected by TRS.

The locations of surface Dirac points in STI and WTI can be deduced from Fig. \ref{fig1}, or by the bulk spectrums with band-inversion character shown in Fig. \ref{fig2}. For cubic lattice, we consider (001) surface, the Dirac points locate at certain TRIM $\mathbf{p}_m$ in 2D BZ, with
$(\mathbf{p}_m,0)$ and $(\mathbf{p}_m,\pi)$ having opposite parities.~\cite{Fu07} From Fig. \ref{fig2}(b) with EHA(I), on can see that the parity $\delta_m$ changes sign between $(\pi,\pi,\pi)$ and $(\pi,\pi,0)$, implying a band inversion at $R$, inducing a single Dirac cone at $(\pi,\pi)$, and this phase is denoted by STI$_{\bar{M}}$ thereafter.
For WTI with EHA(I) in Fig. \ref{fig2}(a), through similar analysis, we expect two Dirac cones at $(\pi,0)$ and $(0,\pi)$, then this phase is denoted by WTI$_{\bar{X}}$. Besides STI$_{\bar{M}}$ with single Dirac cone, in the case of EHA(II), we find a STI$_{\bar{\Gamma} \bar{X}}$ phase with band inversions at three $X$ points (Fig. \ref{fig2}(c)), leading to three Dirac cones, one at $\bar{\Gamma}$, the other two at $\bar{X}$, qualitatively coinciding with the surface states in SmB$_6$.~\cite{Xu14,Yu15,Xu16} For nKI with no band-inversion (Fig. \ref{fig2}(d)), we expect no surface Dirac modes.

In order to verify the surface states in these TI phases, we compute the dispersions on (001) surface. We simulate the cubic lattice with (001) surface by $40$ slabs perpendicular to $z$ axis. With open boundary condition, we write the effective Hamiltonian in $k_x$-$k_y$ space, then obtain the mean-field parameters and chemical potential self-consistently through saddle-point solution, and further diagonalize the Hamiltonian matrix to draw the bulk and surface spectrums. The surface dispersions are displayed in Fig. \ref{fig3} by red solid lines inside the bulk gap, and the locations of Dirac cones confirm above analysis. At half-filling, the Dirac points in WTI$_{\bar{X}}$ and STI$_{\bar{M}}$ all cross the Fermi level (Figs. \ref{fig3}(a) and (b)),
while in STI$_{\bar{\Gamma} \bar{X}}$ (Fig. \ref{fig3}(c)), the Dirac cones form small Fermi rings around $\bar{\Gamma}$ and $\bar{X}$, and the spin texture on these Fermi rings are obtained by calculating spin expectation values, shown in Fig. \ref{fig9}(b). The spin texture shows a helical structure and indicates strong spin-momentum locking in the surface states and reflects the topological nature of STI$_{\bar{\Gamma} \bar{X}}$. Furthermore, In STI$_{\bar{\Gamma} \bar{X}}$ phase, the size of Fermi rings can be enlarged by the difference between $t^\prime_d/t_d$ and $t^\prime_f/t_f$.
The surface states of STI$_{\bar{\Gamma} \bar{X}}$ we demonstrated are in
qualitatively agreement with the surface states of SmB$_6$ derived by previous theoretical calculations~\cite{Lu13,Yu15,Legner15} and SARPES observations,~\cite{Xu13,Xu16} although a model more appropriate than our simplified model is required to describe SmB$_6$.~\cite{Dzero12,Tran12,Yu15}

Evolutions of TI phases with $V$ and $\epsilon_f$ are summarized in Figs. \ref{fig5} (a) and (b), with EHA(I) and EHA(II), respectively, in which the topological transitions are labeled by dashed lines.
Here we should point out that the topological phase boundaries among STI, WTI and nKI through our K-R solutions are very close to Coleman's slave-boson solutions,~\cite{Tran12,Legner14} furthermore, the K-R method has the advantage to include magnetic order conveniently.

In T-PAM, AF order should emerge in the region $\epsilon_f<0$ and weak $V$, thus, once AF order arises in the two phase diagrams in Fig. \ref{fig5}, it should evolve from WTI$_{\bar{X}}$ and STI$_{\bar{\Gamma} \bar{X}}$, respectively. Therefore, EHA(I) and EHA(II) may induce topologically distinct AF states, which will be verified in section V.

\section{$\mathcal{Z}_2$ invariant of AF states in 3D TKI}

Before studying the magnetic transitions in TKI self-consistently, we should derive the expression of topological invariants for these expected AF insulating states by analysing the intrinsic symmetry of the AF Hamiltonian (Eq. (\ref{Hamiltonian})). Firstly, TRS is broken in AF states, because TRS operation $\Theta$ inverts the magnetization at all sites.
Using $\Theta=\textrm{i}I_4\otimes\sigma_y\mathcal{K}$ in present basis (see Eq. (\ref{Hamiltonian})), where $I_4$ denotes $4\times4$ unit matrix and $\mathcal{K}$ is complex conjugation, TRS-breaking in AF phase is manifested by $\Theta\mathbf{H}_{\mathbf{k}}\Theta^{-1}\neq\mathbf{H}_{-\mathbf{k}}$.
 Secondly, space inversion symmetry is preserved in our AF configuration due to existence of inversion center in the middle of a NNN bond: $\mathcal{P}\mathbf{H}_{\mathbf{k}}\mathcal{P}^{-1}=\mathbf{H}_{-\mathbf{k}}$, where the parity matrix is given by $\mathcal{P}=\sigma_z\otimes I_4$.
Breaking of TRS prevents straight-forward application of standard $\mathcal{Z}_2$ topological classification to AF states, so we should find alternative symmetric operation isomorphic to $\Theta$ in AF states.

Since a translation $\mathcal{T}_\mathbf{D}$ by a sublattice vector plus a basic vector of cubic lattice $\mathbf{D}=\mathbf{R}^\prime+\mathbf{a}_i$ (i=1,2,3) inverts the magnetizations of all sites (our AF is staggered between all adjacent sites), the combined operation $\mathcal{S}=\Theta \mathcal{T}_\mathbf{D}$ recoveries the AF configuration, thus the AF states preserve the $\mathcal{S}$-symmetry. The translation operation causes an interchange between two sublattices, so the corresponding operator is $\mathcal{T}_\mathbf{D}(\mathbf{k})=e^{\mathrm{i}\mathbf{k}\cdot\mathbf{D}}I_2\otimes(\sigma_x\otimes I_2)$. It is easy to check the $\mathcal{S}$-symmetry of AF Hamiltonian by $\mathcal{S}_\mathbf{k}\mathbf{H}_\mathbf{k}\mathcal{S}^{-1}_\mathbf{k}=\mathbf{H}_\mathbf{-k}$, with $\mathcal{S}_\mathbf{k}=\Theta \mathcal{T}_\mathbf{D}(\mathbf{k})$=$ e^{-\mathrm{i}\mathbf{k}\cdot\mathbf{D}}I_2\otimes(\sigma_x\otimes \mathrm{i}\sigma_y)\mathcal{K}$.
Secondly, $\mathcal{S}_\mathbf{k}$ is antiunitary like $\Theta$ (because $\mathcal{T}_\mathbf{D}(\mathbf{k})$ is unitary), and $\mathcal{S}^2=\mathcal{S}_\mathbf{k}\mathcal{S}_\mathbf{-k}=-e^{-2\mathrm{i}\mathbf{k}\cdot\mathbf{D}}$,~\cite{Mong10,Fang13} therefore if some of the eight high-symmetry points $\mathbf{k}^\prime_m$ in MBZ satisfy $e^{-2\mathrm{i}\mathbf{k}^\prime_m\cdot\mathbf{D}}=1$, Kramers degeneracy does exist at these points. Due to the antiunitary nature and squaring to minus of $\mathcal{S}$ at these Kramers degenerate momenta (KDM) $\mathbf{k}^\prime_m$, the topology of AF states falls into $\mathcal{Z}_2$ topological class, following directly the $\mathcal{Z}_2$ algebra of TI with inverse symmetry.~\cite{Fu07,Mong10,Fang13} In order to derive the explicit expression of $\mathcal{Z}_2$ invariant for AF states, we should first determine the KDM $\mathbf{k}^\prime_m$.

The AF sublattice is face-centered cubic lattice, and its basic vectors $\mathbf{a}^\prime_i$ and reciprocal basic vectors $\mathbf{b}^\prime_i$ are expressed by unit vectors $\mathbf{x},\mathbf{y},\mathbf{z}$ through
\begin{equation}
\left\{
\begin{array}{lr}
             \mathbf{a}^\prime_1=a(\mathbf{x}+\mathbf{y})\\
             \mathbf{a}^\prime_2=a(\mathbf{y}+\mathbf{z})\\
             \mathbf{a}^\prime_3=a(\mathbf{x}+\mathbf{z})\\
             \end{array}
\right.,
\left\{
\begin{array}{lr}
             \mathbf{b}^\prime_1=\frac{\pi}{a}(\mathbf{x}+\mathbf{y}-\mathbf{z})\\
             \mathbf{b}^\prime_2=\frac{\pi}{a}(-\mathbf{x}+\mathbf{y}+\mathbf{z})\\
             \mathbf{b}^\prime_3=\frac{\pi}{a}(\mathbf{x}-\mathbf{y}+\mathbf{z})\\
             \end{array}
\right.,
\end{equation}
satisfying $\mathbf{a}^\prime_i\cdot\mathbf{b}^\prime_j=2\pi\delta_{ij}$.
The eight high-symmetry points in MBZ are represented by
\begin{align}
\mathbf{k}^\prime_m=\frac{1}{2}(m_1\mathbf{b}^\prime_1+m_2\mathbf{b}^\prime_2+m_3\mathbf{b}^\prime_3),
\end{align}
in which the subscript $m$ represents a combination of $m_i$($i$=1,2,3), with $m_i$ either 1 or 0. Using the translation vector $\mathbf{D}= n_1\mathbf{a}^\prime_1+n_2\mathbf{a}^\prime_2+n_3\mathbf{a}^\prime_3+\mathbf{a}_1$ ($n_1,n_2,n_3$ are integer numbers), we have
\begin{align}
2\mathbf{k}^\prime_m\cdot\mathbf{D}=2\pi\sum_{i=1,2,3}m_in_i+\pi(m_1-m_2+m_3).
\end{align}
To satisfy $e^{-2\mathrm{i}\mathbf{k}^\prime_m\cdot\mathbf{D}}=1$, requires $(m_1-m_2+m_3)=$even, corresponding to four sets of  ($m_1$,$m_2$,$m_3$): (0,0,0), (0,1,1), (1,0,1) and (1,1,0), leading to four KDM: $\Gamma=(0,0,0)$ and three $X$ points $(\pi,0,0)$, $(0,\pi,0)$, $(0,0,\pi)$, which are just four out of eight TRIM in PM phases. The other four high-symmetry points in MBZ are not KDM thus are irrelevant to determine the topological invariant.

At four KDM $\mathbf{k}^\prime_m$, $\mathbf{H}_{-\mathbf{k}^\prime_m}$=$\mathbf{H}_{\mathbf{k}^\prime_m}$, the inverse symmetry ensures the commutation relation $[\mathbf{H}_{\mathbf{k}^\prime_m},\mathcal{P}]=0$, hence the eigenstates at KDM are also parity eigenstates with parities $\pm1$.
The $\mathcal{S}$ and $\mathcal{P}$ symmetries lead to a new topological classification of AF states by the $\mathcal{Z}_2$ invariant,~\cite{Mong10,Fang13} determined by the quantities $\delta^\prime_m$ at four KDM $\mathbf{k}^\prime_m$,
which are calculated by the parities at $\mathbf{k}^\prime_m$ through
\begin{align}
\delta^\prime_m=\prod_i\xi_i(\mathbf{k}^\prime_m),\label{deltaprime}
\end{align}
where $\xi_i(\mathbf{k}^\prime_m)$ is the parity of the occupied state $i$ at $\mathbf{k}^\prime_m$, and each Kramers pair is multiplied only once in $\delta^\prime_m$.
Similar to the PM phases in TKI, effective hybridization in AF states also vanishes at KDM, then the AF dispersions $E^{(i)}_{\mathbf{k}^\prime_m}$ equal either $\epsilon^d_{\mathbf{k}^\prime_m}$ or $\tilde{\epsilon}^f_{\mathbf{k}^\prime_m}$, so the parity is determined by
\begin{equation}
\xi_i(\mathbf{k}^\prime_m)=\left\{
             \begin{array}{lr}
             1,\mathrm{if}E^{(i)}_{\mathbf{k}^\prime_m}=\epsilon^d_{\mathbf{k}^\prime_m}\\
             -1,\mathrm{if}E^{(i)}_{\mathbf{k}^\prime_m}=\tilde{\epsilon}^f_{\mathbf{k}^\prime_m}\\
             \end{array}
\right..
\end{equation}
The $\mathcal{Z}_2$ invariant $\nu^\prime$ in AF state is then defined by
\begin{align}(-1)^{\nu^\prime}=\prod_{\mathbf{k}^\prime_m\in \mathrm{KDM}}\delta^\prime_m.\label{num}
\end{align}
$\nu^\prime$ is the only topological index in AF states, and it is strongly related to the strong topological index $\nu_0$ in PM TI phases. Near the AF transition point, an infinitesimal AF order arises, then the 3D BZ is folded into MBZ, and the eight TRIM in PM phases are also folded into four KDM in AF phases, explicitly, $R$ and three $M$ are folded into $\Gamma$ and three $X$ points, respectively. In addition, the spectrums are also folded, leading to two occupied dispersions under Fermi level, then the quantity $\delta^\prime_m$ at a KDM $\mathbf{k}^\prime_m$ is essential the product of two $\delta_m$ of two corresponding TRIM which are folded into $\mathbf{k}^\prime_m$, therefore the product of parities at four KDM in AF state equals the product of parities at all eight TRIM in PM phase, in this sense, near the magnetic boundaries, $\nu^\prime$ of AF phase is equivalent to strong topological index $\nu_0$ of PM phase from which AF order grows.
Then we can draw to a conclusion that an AF order growing from STI (by reducing $V$, etc) leads to an AFTI with $\nu^\prime=1$; if AF order is induced from WTI or nKI, a nAFI arises with $\nu^\prime=0$, which helps us to search AF states with different topologies. When leaving the magnetic boundaries, AF magnetization increases, then $\nu^\prime$ should be calculated by Eq. (\ref{num}), and it may be shifted by varying model parameters to arouse a topological transition between AF states, which is to be discussed in the following sections.

The topologies of AF states will be reflected by the surface states. On AF-ordered (001) surface, the $\mathcal{S}$ symmetry is conserved, with vector $\mathbf{D}$ parallel to the surface, then the surface MBZ has two KDM: $\bar{\Gamma}=(0,0)$ and $\bar{X}=(0,\pi)$. As AF vector $\mathbf{Q}=(\pi,\pi)$ on this surface, $\bar{M}=(\pi,\pi)$ is equivalent to $\bar{\Gamma}$ due to folding of 2D BZ, and another $\bar{X}$ point $(\pi,0)$ is equivalent to $(0,\pi)$, therefore the four TRIM ($\bar{\Gamma}$, $\bar{M}$ and two $\bar{X}$) in surface BZ of PM phases now become $\mathcal{S}$-invariant momenta, and they are KDM which may support gapless Dirac modes.~\cite{Fang13} Analogous to STI, on the AF-ordered surfaces in AFTI, on which $\mathcal{S}$-symmetry is preserved, there are odd number of gapless Dirac cones at certain KDM which are robust against $\mathcal{S}$-preserving perturbations. We remind that the topological classification requires insulating AF states with full bulk gap,~\cite{Mong10,Fang13} otherwise the topological argument will be meaningless.

\begin{figure}[tbp]
\hspace{-0cm} \includegraphics[totalheight=2.47in]{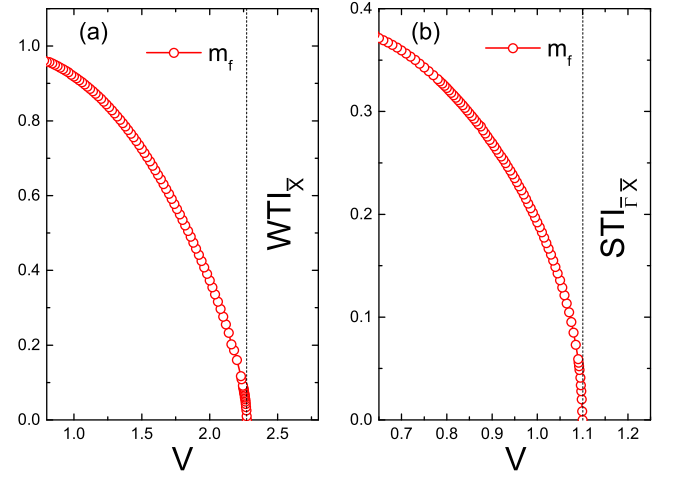}
\caption{(Color online)
Evolution of staggered magnetization $m_f$ with hybridization strength $V$, indicating a second-order magnetic transition.
Model parameters: (a) $\epsilon_f=-3$ with EHA(I), (b) $\epsilon_f=-1$ with EHA(II). For (a) and (b), the AF phases are in the vicinity of WTI$_{\bar{X}}$ and STI$_{\bar{\Gamma} \bar{X}}$, respectively.
}
\label{fig4}
\end{figure}

\begin{figure}[tbp]
\hspace{-0.4cm} \includegraphics[totalheight=1.8in]{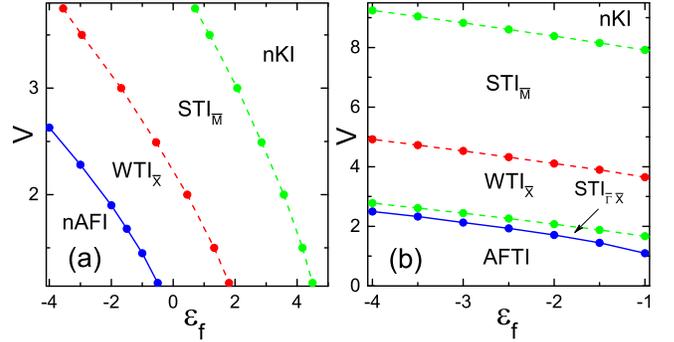}
\caption{(Color online)
Phase diagrams on $\epsilon_f$-$V$ plane, showing magnetic transitions (solid blue lines) and topological transitions among TI phases (dashed lines). EHA(I) for (a) and EHA(II) for (b). Depending on whether the AF order arises from WTI or STI, AF phases are classified into nAFI or AFTI, respectively.
}
\label{fig5}
\end{figure}

\begin{figure}[tbp]
\hspace{-0.3cm} \includegraphics[totalheight=2.67in]{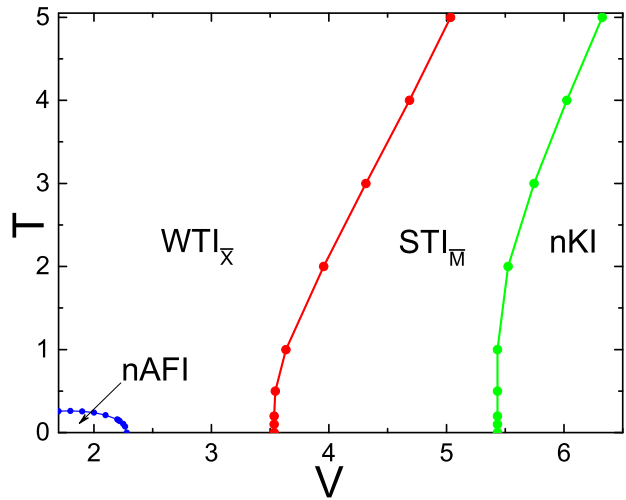}
\caption{(Color online)
Phase evolution with hybridization strength $V$ and temperature $T$. Parameters: $\epsilon_f=-3$ and EHA(I).
}
\label{fig6}
\end{figure}

\heavyrulewidth=1bp

\begin{table*}
\small
\renewcommand\arraystretch{1.3}
%\centering
\caption{\label{table1}
$\mathcal{Z}_2$ classification and surface Dirac points of the AF states
demonstrated in Fig. \ref{fig7}. The last two columns show critical
hybridization of the magnetic transitions under EHA and $\epsilon_f$
 in the second column, and topologies of the PM phases near the magnetic transitions. }
%\begin{ruledtabular}
\begin{tabular*}{17cm}{@{\extracolsep{\fill}}ccccccccc}
\toprule
    classification & model parameters & dispersions\footnotemark[1]& $\delta^\prime_m$ at $\Gamma$ & $\delta^\prime_m$ at three $X$ & $\nu^\prime$\footnotemark[2]  &  Dirac points\footnotemark[3] & critical $V$ & PM phase   \\
\hline
 nAFI & EHA(I), $\epsilon_f=-3$, $V=2$& Fig. \ref{fig7}(b), \ref{fig7}(a)&  -1 & -1 & 0 & - &2.27 & WTI$_{\bar{X}}$ \\
 AFTI & EHA(II), $\epsilon_f=-2$, $V=1.6$ &  Fig. \ref{fig7}(d), \ref{fig7}(c) &1 & -1 & 1  & $\bar{\Gamma}$ and $\bar{M}$ & 1.71 & STI$_{\bar{\Gamma}\bar{X}}$\\
   nAFI & $\epsilon_f=-2$, $V=2.05$\footnotemark[4]  & Fig. \ref{fig7}(f), \ref{fig7}(e) & -1 & -1 & 0 &  - & 2.22 & WTI$_{\bar{X}}$ \\
\bottomrule
\end{tabular*}
%\end{ruledtabular}
\footnotetext[1]{The surface dispersions are showing on (001) surface.}
\footnotetext[2]{The $\mathcal{Z}_2$ index $\nu^\prime$ is calculated by Eq. (\ref{num}). }
\footnotetext[3]{Showing topologically protected Dirac points in the AF phases}
\footnotetext[4]{With EHA: $t^{\prime}_d=t^{\prime\prime}_d=-0.375$, $t_f=-0.1$, $t^\prime_f=t^{\prime\prime}_f=0.0375$.}
\label{data}
\end{table*}

\section{magnetic transitions, $\mathcal{Z}_2$ classification of AF phases, and the surface states}

Based on the topological phase diagrams of TI phases in Fig. \ref{fig5}, we now turn to the magnetic transitions in TKI. We perform a saddle point solution for the ground-state energy of AF state (Eq. (\ref{FAF})) to determine the mean-field parameters $n_f$, $m_f$, $h$, $\eta$ and the chemical potential $\mu$, and obtain the bulk dispersions by diagonalizing the Hamiltonian matrix (Eq. (\ref{Hk})), then further calculate $\nu^\prime$ by Eq. (\ref{num}) to classify the solved AF phases.

\begin{figure}[tbp]
\hspace{-0.2cm} \includegraphics[totalheight=3.265in]{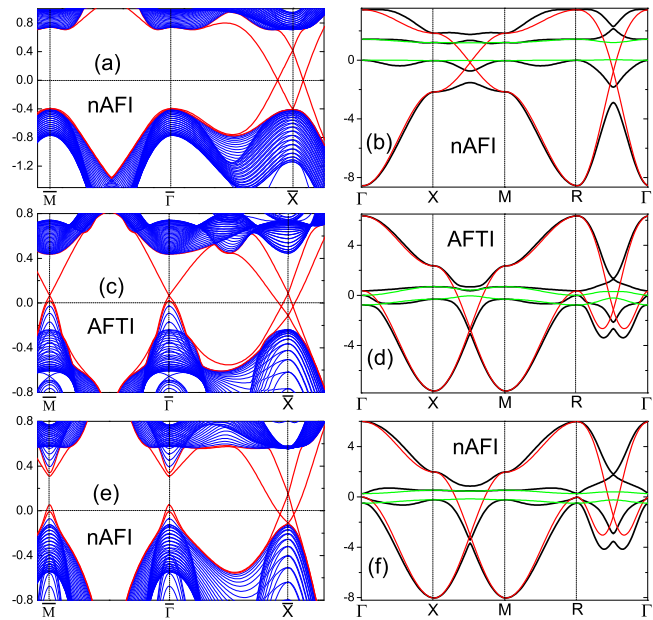}
\caption{(Color online)
Left column: surface dispersions (red lines) on (001) surface through slab-calculations. Right column: corresponding 3D bulk spectrums comparing with $d$- (red lines) and renormalized $f$- dispersions (green lines). The upper and down rows are for nAFI states, while the middle row is for AFTI. Parameters: (a) and (b) with $\epsilon_f=-3$, $V=2$ and EHA(I); (c) and (d) with $\epsilon_f=-2$, $V=1.6$ and EHA(II); (e) and (f) with $\epsilon_f=-2$, $V=2.05$, $t^{\prime}_d=t^{\prime\prime}_d=-0.375$, $t_f=-0.1$, $t^\prime_f=t^{\prime\prime}_f=0.0375$.
}
\label{fig7}
\end{figure}

In Figs. \ref{fig4}(a) and (b), we display the calculated order parameters $m_f$ and $h$ as functions of $V$, for EHA(I) at fixed $\epsilon_f=-3$, and for EHA(II) at $\epsilon_f=-1$, respectively. The critical behavior of $m_f$ with respect to $V$ clearly indicates a second-order magnetic transition.
In Figs. \ref{fig5}(a) and (b), the critical hybridization $V_c$ of magnetic transition is plotted as $\epsilon_f$ varies.
The suppression of $V_c$ as $\epsilon_f$ approaches the Fermi level is attributed to the enhancement of valence fluctuation of $f$ electrons which suppresses the AF order.

The phase diagrams are then summarized on $\epsilon_f$-$V$ plane in Figs. \ref{fig5}(a) and (b), for EHA(I) and EHA(II), respectively, including both AF phases and TI phases. For EHA(I), we find that the AF phase is in proximity to WTI, indicating a nAFI with $\nu^\prime=0$; while for EHA(II), the AF phase is in proximity to STI, indicating an AFTI with $\nu^\prime=1$. The bulk dispersions of three typical AF states (two for nAF and one for AFTI) are demonstrated in Figs. \ref{fig7}(b), (d) and (f), through which we can calculate $\delta^\prime_m$ at four KDM using Eq. (\ref{deltaprime}), then obtain $\nu^\prime$ by Eq. (\ref{num}). The results are shown in Tab. \ref{table1}, confirming our analysis.

In Fig. \ref{fig6}, we display the phase evolution with $V$ and temperature $T$ at fixed $\epsilon_f=-3$ with EHA(I). It shows that while $V$ is enhanced, the N\'{e}el temperature of nAFI phase is suppressed continuously, then vanishes at $V_c$, which is the critical value of magnetic transition at zero temperature. Therefore, besides by the enhancement of $V$ or ascent of $\epsilon_f$ at zero temperature, the nAFI-WTI transition (and AFTI-STI transition) can also be driven by increasing temperature. In addition, the phase boundaries among WTI, STI and nKI are slightly shifted by finite temperatures.

\begin{figure}[tbp]
\hspace{-0.2cm} \includegraphics[totalheight=2.58in]{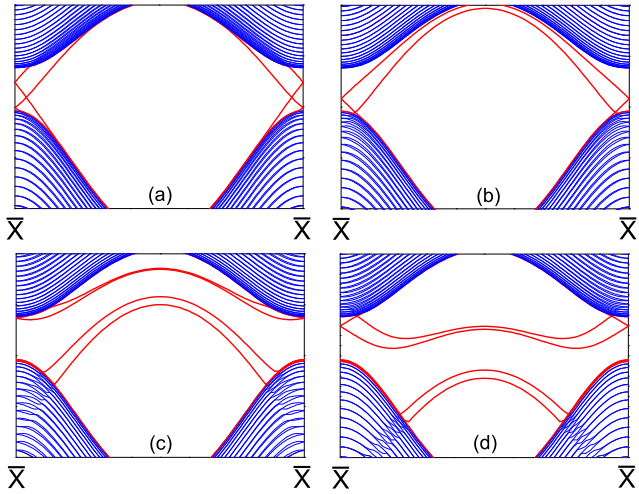}
\caption{(Color online)
Under continuously enhanced contact potential on (001) surface from (a) to (d),
the surface dispersions (red lines) in nAFI are deformed adiabatically to open a full gap in (d), indicating the trivial nature of the surface states in nAFI. The surface states around $\bar{X}$ in AFTI exhibit similar feature and are also topologically trivial.
}
\label{fig8}
\end{figure}

Now we calculate the surface states of AF phases. With three sets of model parameters listed in Tab. \ref{table1} (two for nAFI with bulk spectrums displayed in Figs. \ref{fig7}(b) and (f), and one for AFTI shown in Fig. \ref{fig7}(d)),
we have performed saddle-point solutions of 40 slabs to derive the mean-field parameters and chemical potential, then diagonalize the Hamiltonian matrix to obtain the surface dispersions on (001) surface. The bulk and surface dispersions are given in Figs. \ref{fig7}(a) and (e) for nAFI, and (c) for AFTI, in which the bulk spectrums are all full-gapped.
For nAFI, the surface states remain gapless within present solution. The original Dirac cone at $\bar{X}$ in WTI$_{\bar{X}}$ (see Fig. \ref{fig3}(a)) is decomposed upwards and downwards into two Dirac cones by AF magnetization, leading to two Dirac points at $\bar{X}$, one above and another below the Fermi level, similar to that reported in Ref. \onlinecite{Chang17}.
Since $\bar{X}$ is now KDM, therefore, in nAFI, Kramers degeneracy takes place at these two Dirac points on both sides of the Fermi level. However, such surface states cross the Fermi level even times from $\bar{X}$ to any other KDM, revealing the non-topologically-protected nature of nAFI, in the manner that these surface states can be deformed adiabatically by perturbations which conserve $\mathcal{S}$ symmetry (e.g., adding contact potential on surface),~\cite{Fang13} to gap the surface dispersions,~\cite{Fu07} see Fig. \ref{fig8}. Hence, although the surface states in nAFI remain gapless by present solution, nAFI is topologically undistinguishable from other AF states with gapped surface states (such as adding an AF order to nKI).
AFTI shown in Fig. \ref{fig7}(c) is in the vicinity of STI$_{\bar{\Gamma}\bar{X}}$ phase, and the Dirac point at $\bar{\Gamma}$ persists, because $\Gamma$ point shares opposite $\delta^\prime_m$ with other three KDM (see Tab. \ref{table1}). Another Dirac point at $\bar{M}$ is formed by folding of surface BZ, and these two Dirac cones in AFTI are topologically protected against $\mathcal{S}$-conserving perturbations. In addition, the original Dirac cones at $\bar{X}$ are decomposed in AFTI, and the surface states around $\bar{X}$ in AFTI are also topologically trivial, similar to nAFI.

\begin{figure}[tbp]
\hspace{-0.2cm} \includegraphics[totalheight=1.72in]{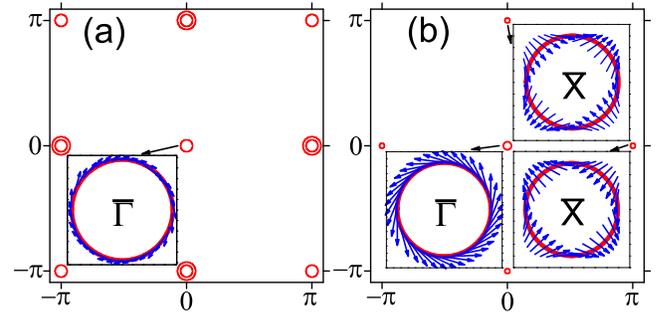}
\caption{(Color online)
 Fermi rings (red circles) and spin textures (blue arrows) on (001) surface, showing on $k_x$-$k_y$ plane.
(a) Fermi rings in AFTI, inset shows the helical spin texture on the Fermi ring around $\bar{\Gamma}$.
(b) Fermi rings and helical spin texture in STI$_{\bar{\Gamma} \bar{X}}$. Parameters in AFTI are set equal to Fig. \ref{fig7} (c), while for STI$_{\bar{\Gamma} \bar{X}}$, parameters are chosen according to Fig. \ref{fig3} (c). Although the surface states around $\bar{X}$ in AFTI remain gapless within present solution, they are trivial and can be gapped adiabatically, while the surface states around $\bar{\Gamma}$ and $\bar{M}$ in AFTI are topologically protected.
}
\label{fig9}
\end{figure}

Although bulk gap persists, the gapless surface states in AFTI induce metallic Fermi rings, as shown in Fig. \ref{fig9}(a), comparing with the Fermi rings in STI$_{\bar{\Gamma}\bar{X}}$ shown in Fig. \ref{fig9}(b). For AFTI, the Fermi ring around $\bar{\Gamma}$ evolves continuously to that in STI$_{\bar{\Gamma}\bar{X}}$, when approaching the magnetic boundary.
It should be noted that since the surface states around $\bar{X}$ in AFTI are topologically trivial, corresponding Fermi surfaces around $\bar{X}$ can be destructed by $\mathcal{S}$-conserving perturbations such as contact potential on the surfaces; on the contrary, the Fermi rings around $\bar{\Gamma}$ and $\bar{M}$ in AFTI are topologically protected and are robust under $\mathcal{S}$-conserving perturbations. The spin texture on the Fermi ring around $\bar{\Gamma}$ in AFTI is illustrated by the inset of Fig. \ref{fig9}(a), showing a helical spin structure and indicating spin-momentum locking in the surface Dirac states, however, on (001) surface, large N\'{e}el energy suppresses the projection of spins on this surface, therefore, the spin-momentum locking in AFTI is much weaker than in STI$_{\bar{\Gamma}\bar{X}}$ (Fig. \ref{fig9}(b)). On the Fermi rings around $\bar{X}$ points in AFTI, we found that the spins are greatly suppressed, and no clear signal of spin-momentum locking is observed.

As revealed by previous theoretical calculations~\cite{Lu13,Yu15,Legner15} and SARPES observations,~\cite{Xu13,Xu16} the TKI compound SmB$_6$ is in a STI$_{\bar{\Gamma}\bar{X}}$ phase, hence the pressure-induced magnetic phase in SmB$_6$ is most probably an AFTI, but under two conditions: firstly, no former topological transition takes place before the magnetic transition, secondly, there is a lattice translation which flips the magnetization at all sites, allowing the application of $\mathcal{Z}_2$ classification to the AF phase.
However, during the high-pressure-induced magnetic transition, tracking the evolution of model parameters is a complicated task, so the existence of AFTI in SmB$_6$ requires further first-principle investigations and experimental confirmation.

In an AFTI, the surface states are strongly anisotropic to surface orientation.~\cite{Fang13} AF-ordered (001) surface we studied above conserves the $\mathcal{S}$ symmetry, leading to topologically protected gapless Dirac cones on it, namely these surface states are stable under additional interactions which do not violate $\mathcal{S}$ symmetry. On ferromagnetically (FM) ordered surfaces which violate $\mathcal{S}$ symmetry (e.g., (111)surface), no KDM exists, so the surface states are generally gapped,~\cite{Mong10,Fang13} similar to the surface states in magnetic topological insulators.~\cite{Chang13}

The topologies of the AF phases depend on the topologies of the PM phases from which the AF orders grow, therefore, the AFTI phase only emerges below the critical $V_c$ of a parameter region in which STI phase is favored. Consequently, AFTI only emerges in a narrow parameter region.
The nAFI and AFTI states we derived are close to WTI$_{\bar{X}}$ and STI$_{\bar{\Gamma}\bar{X}}$, respectively.
From Fig. \ref{fig5}, one can see that for physical $\epsilon_f<0$, STI$_{\bar{M}}$ phase emerges at relative strong $V$ at which magnetic order has already been suppressed. By adding further EHA properly, we expect to realize an AF transition inside STI$_{\bar{M}}$ phase, then the arising AF state is an AFTI, with protected Dirac points at $\bar{\Gamma}$ and $\bar{M}$ on (001) surface, but no gapless state at $\bar{X}$ even within present solution, which is different from AFTI shown in Fig. \ref{fig7}(c).

\section{topological transition between AF states}

In the above, we have shown that with EHA(I) or EHA(II), a magnetic transition takes place to induce nAFI or AFTI, respectively. Particulary, in Fig. \ref{fig5}(b), one can see that with EHA(II), although the magnetic transition occurs between STI$_{\bar{\Gamma}\bar{X}}$ and AFTI, the magnetic boundary is quite close to STI$_{\bar{\Gamma}\bar{X}}$$-$WTI$_{\bar{X}}$ boundary. On the other hand, we find that the critical $V_c$ of magnetic transition increases rapidly with decreasing $|t_f|$ (note $t_f<0$ to get an insulating bulk state), on the contrary, $V_c$ of STI$_{\bar{\Gamma}\bar{X}}$$-$WTI$_{\bar{X}}$ transition decreases as $|t_f|$ is reduced. Consequently, when $|t_f|$ is less than a critical value, the magnetic transition now occurs nearby WTI$_{\bar{X}}$ to induce a nAFI phase. To see this, we fixe $\epsilon_f=-2$, shift $t_f$ in EHA(II) to $-0.1$ and keep the relation $t^\prime_f=t^{\prime\prime}_f=-0.525t_f-0.015$ which EHA(II) obeys, then find a magnetic transition at $V_c=2.22$, now from WTI$_{\bar{X}}$ to nAFI state (see top right corner on the red line in Fig. \ref{fig11}). Therefore, a continuous change of $t_f$ (and associated change of $t^\prime_f$ and $t^{\prime\prime}_f$) can shift the magnetic boundary from nearby STI$_{\bar{\Gamma}\bar{X}}$ to nearby WTI$_{\bar{X}}$, shown by the two segments of red line in Fig. \ref{fig11}.
From knowledge of the $\mathcal{Z}_2$ classification of AF states discussed in section IV, upon reduction of $V$ from these two segments of magnetic boundaries, we obtain an AFTI and nAFI states, respectively. Therefore, by varying $t_f$ in such way, we can realize a topological transition between AFTI and nAFI.

\begin{figure}[tbp]
\hspace{-0.2cm} \includegraphics[totalheight=2.65in]{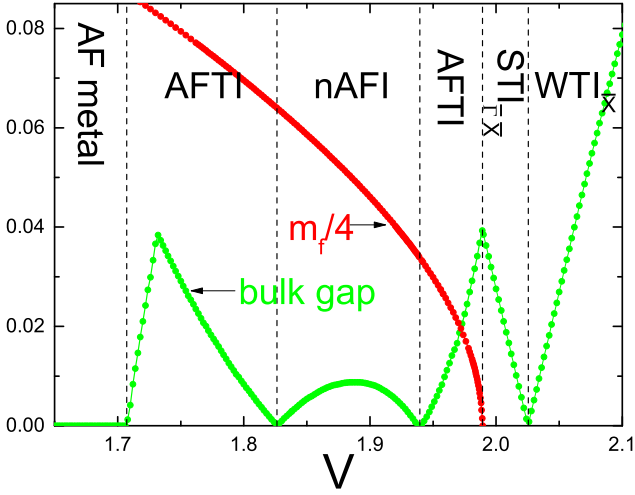}
\caption{(Color online)
Bulk gap and $m_f$ vs $V$, showing the magnetic transition, the topological transitions, and the insulator-metal transition. The topological transitions between AFTI and nAFI are driven by closing of bulk gap.
Model parameters: $\epsilon_f=-2$, $t^\prime_d=t^{\prime\prime}_d=-0.375$, $t_f=-0.144$, $t^\prime_f=t^{\prime\prime}_f=0.0606$.
}
\label{fig10}
\end{figure}

To verify this topological transition, we set $t_f$=$-0.144$ and $\epsilon_f$=$-2$ to calculate the phase evolution with $V$, the result is illustrated in Fig. \ref{fig10}. Magnetic transition occurs between AFTI and STI$_{\bar{\Gamma}\bar{X}}$ at $V=1.989$. With decreasing $V$, we find two successive topological transitions between AF phases: an AFTI-nAFI transition at $V=1.941$ and a following nAFI-AFTI transition at $V=1.826$. At these two topological transitions, the bulk gap is closed, while at the magnetic transition, bulk gap persists. When further reducing $V$ after nAFI-AFTI transition, bulk gap increases then decreases rapidly to generate an insulator-metal transition at $V=1.707$, and the metallic AF state is similar to the AF states proposed in Ref. \onlinecite{Chang17,ZhiLi15}. Phase diagram summarizing the magnetic transitions, topological transitions and insulator-metal transition are shown in Fig. \ref{fig11}.

To see explicitly what happens during the topological transition between AFTI and nAFI, we depicted the bulk and surface dispersions of the two phases on both sides of this transition in Figs. \ref{fig7}(c)-(f). We find that AFTI-nAFI transition is accompanied by closing of bulk gap at $\Gamma$ and equivalent $R$ points. On AFTI side of this transition (Figs. \ref{fig7}(c) and (d)), $\delta^\prime_m=1$ at $\Gamma$, while at three $X$ points, $\delta^\prime_m=-1$, inducing a topological index $\nu^\prime=1$. By closing and reopening of the bulk gap at $\Gamma$ (and $R$) during AFTI-nAFI transition, in nAFI phase, $\delta^\prime_m$ is shifted to $-1$ at $\Gamma$ and remains $-1$ at $X$, leading to $\nu^\prime=0$. So gap-closing during AFTI-nAFI transition causes a parity inversion at $\Gamma$, shifts the topological index, consequently leads to the vanishing of Dirac cones at $\bar{\Gamma}$ and $\bar{M}$ in nAFI.

\begin{figure}[tbp]
\hspace{-0cm} \includegraphics[totalheight=2.62in]{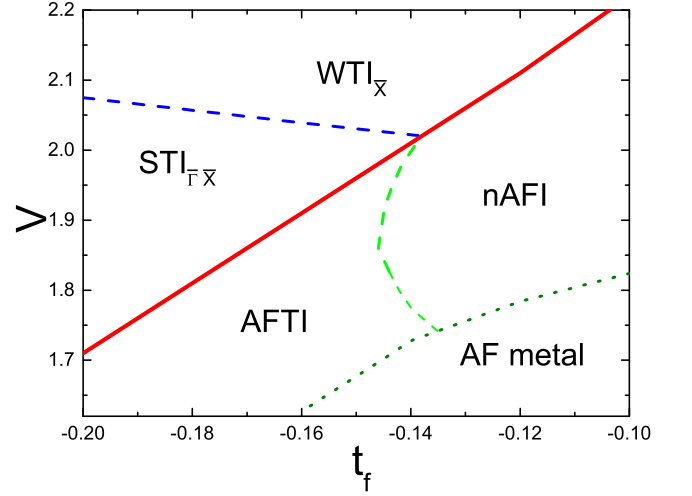}
\caption{(Color online)
Phase evolution with $t_f$ and $V$. The red solid line indicates magnetic boundary, the dashed lines denote topological transitions, while the dotted line stands for insulator-metal transition. Model parameters: $\epsilon_f=-2$, $t^\prime_d=t^{\prime\prime}_d=-0.375$, $t^\prime_f=t^{\prime\prime}_f=-0.525t_f-0.015$.
}
\label{fig11}
\end{figure}

\heavyrulewidth=1.bp
\begin{table*}
\small
\renewcommand\arraystretch{1.3}
%\centering
\caption{\label{table3}
 Comparison between our work and some similar works on TKI.}
%\begin{ruledtabular}
%\begin{tabular*}{17cm}{@{\extracolsep{\fill}}ccccccc}
\begin{tabular}{m{2cm}<{\centering}m{1.3cm}<{\centering}m{1.7cm}<{\centering}m{2.3cm}<{\centering}m{2.1cm}<{\centering}m{2.2cm}<{\centering}m{2.6cm}<{\centering}m{2.6cm}<{\centering}}
\toprule
     Group & Model & Hopping  & Parameter regime & Solving method  & Symmetry   & Phases &  Transitions      \\
\hline

Present authors (2018)   &  T-PAM\footnotemark[1] & NN, NNN, NNNN    &   Half-filling, infinite $U$, insulating   &   Slave boson   &   $\mathcal{S}$ symmetry  &   STI, WTI, nKI, AFTI, nAFI& Among TI$_s$\footnotemark[3], magnetic transition, AFTI-nAFI  \\\hline
Peters et al. (2018)\cite{Peters18}   &   T-PAM\footnotemark[1] & NN, NNN, NNNN   &   Off half-filling, large $U$   &   DMFT   &  Reflection symmetry of cubic lattice   &   FM, chiral surface states& Magnetic transition \\\hline
Chang \& Chen (2018)\cite{Chang17}  & -& -  &  Small $U$, metallic   &    First-principle  &   $\mathcal{S}$ symmetry   &  Topological AF&Pressure-induced magnetism  \\\hline
Legner et al. (2014)\cite{Legner14} &  T-PAM\footnotemark[1] & NN, NNN, NNNN    &  Finite $U$, insulating   & Slave boson & TRS & STI, WTI, nKI& Among TI$_s$\footnotemark[3] \\\hline
Tran el al. (2012)\cite{Tran12}    &  T-PAM\footnotemark[2] & NN  & Infinite $U$, insulating  & Slave boson &  TRS  &  STI, WTI, nKI& Among TI$_s$\footnotemark[3]\\
\bottomrule
%\end{tabular*}
\end{tabular}
%\end{ruledtabular}
\footnotetext[1]{The simplified T-PAM, see Eq. (\ref{PAM}).}
\footnotetext[2]{The T-PAM with $s$ and $f$ orbits.   }
\footnotetext[3]{Topological transitions among STI, WTI and nKI, including STI-WTI, WTI-nKI and STI-nKI transitions.  }
\label{data}
\end{table*}

In this work, the nAFI-AFTI topological transition is achieved by special setting and variation of some model parameters. When applying pressure to real TKI materials, the variation of model parameters can be quite complicated to track, so whether such nAFI-AFTI topological transition is realizable deserves further first-principle calculations and experimental verification. Besides, quantum phase transitions between topological trivial and nontrivial AF phases and the metal-insulator transition in other system have been reported in
literature.~\cite{Baireuther14,Kimura18}

\section{conclusion and discussion}

In conclusion, we have verified a novel topologically protected AFTI phase in 3D TKI, and realized a topological transition between AFTI and nAFI, for the first time. We have performed an extended slave-boson mean-field solution of 3D TKI modeling by the half-filled periodic Anderson model with spin-orbit coupled $d$-$f$ hybridization in large $U$ limit.
In a wide parameter region, we have found second-order transitions from TI phases to AF phases. Although time-reversal symmetry $\Theta$ is broken, the AF phases preserve the symmetry under combined operation $\mathcal{S}=\Theta T_{\mathbf{D}}$, in which a translation by vector $\mathbf{D}$ flips the magnetization at all sites. $\mathcal{S}$ operator is antiunitary and squares to minus at four out of eight high-symmetry points in MBZ, leading to Kramers degeneracy at these momenta. The Kramers degeneracy at four KDM and the inverse symmetry of AF result in a new type of $\mathcal{Z}_2$ classification for AF states by invariant $\nu^\prime$ calculated by product of parities at four KDM, and $\nu^\prime$ is in analogy to the STI index $\nu_0$. By applying $\mathcal{Z}_2$ classification to the slave-boson solutions of AF states, we found two topologically distinguishable AF phases in different parameter regions, one is AFTI with topologically protected gapless Dirac cones around $\bar{\Gamma}$ and $\bar{M}$ on (001) surface, exhibiting helical spin texture; the other is nAFI with trivial surface states, and the magnetic transition occurs either between AFTI and STI$_{\bar{\Gamma}\bar{X}}$, or between nAFI and WTI$_{\bar{X}}$.
Under special variations of model parameters, we observed a topological transition between nAFI and AFTI driving by closing of bulk gap, and an insulator-metal transition inside AF phases. The magnetic transitions, the topological transition between nAFI and AFTI, and the topological transitions between TI phases have been summarized in a global phase diagram. We should emphasize that our derived AFTI in 3D TKI is an insulator with full bulk gap, distinct from the AF phases with metallic bulk previously reported in SmB$_6$ and GdBiPt by other authors.~\cite{Chang17,ZhiLi15}

 In this paper, we have so far discussed an AF structure which is staggered between adjacent sites. Actually, other AF configurations have also been discussed in literature, e.g., the so-called A-AF state which is staggered along one axis.~\cite{Chang17} We have pointed out that as long as an insulating AF phase possesses a translation by a certain lattice vector which flips the magnetization at all sites, plus it has an inversion center, then this AF state can be classified by the $\mathcal{Z}_2$ invariant calculated by the parities at four KDM in its MBZ. Under these conditions, we can always reach the conclusion that if an insulating AF phase evolves from STI, it is an AFTI; while it arises from WTI, it is nAFI, regardless of the detailed AF configurations. Therefore, once an insulating A-AF state emerges (in different parameter regions from our work), it can also be classified through our scheme.
The present topological classification for AF states is based on the $\mathcal{S}$-symmetry, actually, for AF states possess other symmetry, e.g., mirror symmetry or reflection symmetry, AF states can be classified by different algebras.~\cite{Peters18,Kimura18} In Tab. \ref{table3}, we present a brief description of some related works on TKI in literature and compare them with our work.

 Our slab-simulations have used an uniform solution of the mean-filed parameters, which may actually vary near the surface and may cause interesting effects such as surface Kondo breakdown and light surface states.~\cite{Alexandrov15,Peters16,Erten16} Surface Kondo breakdown or decoupling between $d$-$f$ electrons on surface (and consequently location of $f$ electrons or $n_f$$\rightarrow1$) is ascribed to $b$$\rightarrow0$ or vanishing of effective Kondo hybridization on surface of TKI.~\cite{Alexandrov15} In our algebra, using $Z=\sqrt{2(1-n_f)/(2-n_f)}$, one can see that $n_f$$\rightarrow1$ on surface leads naturally to vanishing of effective hybridization $VZ$, in this sense, $Z$ is analogous to $b$, therefore, surface Kondo breakdown can also be treated by K-R method. How the AF states behave near surface and whether surface Kondo breakdown takes place in AF phases and the consequence to the surface dispersions, remains an open issue.
We look forward to obtain the surface dispersions using depth-dependent solutions, to deeply investigate the surfaces states in AFTI.
The widely used slave-boson approach has provided a rather satisfactory description of TKI in terms of topology and topological transitions. The limitation of this mean-filed approach lies in low efficiency to describe the dynamic behaviors such as Kondo resonance, Kondo screening and spin or charge correlations, which may require more rigorous methods such as Monte Carlo simulation.~\cite{Vekic95}

\acknowledgments

H. Li is supported by NSFC (No.
11764010, 11504061) and Guangxi Natural Science Foundation (No. 2017GXNSFAA198169, 2015GXNSFBA139010).
Y. Zhong is supported by NSFC (No. 11704166) and the Fundamental Research Funds for the Central Universities.
Y. Liu thanks SPC-Lab Research Fund (No. XKFZ201605).
H. F. Song thanks the supports by Science Challenge Project (No. TZ2018002), NSFC (No. 11176002), and Foundation of LCP (No. 6142A05030204).

\begin{appendices}

\section{the saddle-point equations for AF phases}

  Here we discuss the equations for saddle point solution of AF phase based on the mean-field Hamiltonian Eq. (\ref{Hamiltonian}). Since the Hamiltonian matrix $\mathbf{H}_{\mathbf{k}}$ (Eq. (\ref{Hk})) has no analytical expression for its eigenvalues, we have to solve it numerically to get the unitary transformation matrix $\mathrm{U}_{\mathbf{k}}$ which diagonalize $\mathbf{H}_{\mathbf{k}}$: $\mathrm{U}^\dag_{\mathbf{k}}\mathbf{H}_{\mathbf{k}}\mathrm{U}_{\mathbf{k}}=\Lambda_{\mathbf{k}}$,
or $\Psi^\dag_{\mathbf{k}}\mathbf{H}_{\mathbf{k}}\Psi_{\mathbf{k}}=\Phi^\dag_{\mathbf{k}}\Lambda_{\mathbf{k}}\Phi_{\mathbf{k}}$,
where $\Lambda_{\mathbf{k}}$ is a diagonal matrix with the eigenvalues $E^{(i)}_{\mathbf{k}} (i=1,...,8)$ as its diagonal elements, $\Phi^\dag_{\mathbf{k}}$ is the eight-component creation operator for elementary excitations. Now we can calculate the ground-state expectation values for quadratic combinations of $d$ and $f$ operators ($n,m=1,...,8$):
\begin{align}\langle nm\rangle_{\mathbf{k}}\equiv&\langle (\Psi^\dag_{\mathbf{k}})_n(\Psi_{\mathbf{k}})_m\rangle\nonumber\\
=&\langle\sum_i(\Phi^\dag_{\mathbf{k}})_i(\mathrm{U}_{\mathbf{k}})^*_{ni}\sum_j(\mathrm{U}_{\mathbf{k}})_{mj}(\Phi_{\mathbf{k}})_j\rangle\nonumber\\
=&\sum^8_{i=1}(\mathrm{U}_{\mathbf{k}})^*_{ni}(\mathrm{U}_{\mathbf{k}})_{mi}\Theta(-E^{(i)}_{\mathbf{k}}).
\end{align}
Using these expectation values $\langle nm\rangle_{\mathbf{k}}$ which can be extracted from numerical diagonalization, we can write the free energy $F$ as the expectation value of the Hamiltonian Eq. (\ref{Hamiltonian}):
\begin{align}F=N(hm_f-\eta n_f)+\sum_{\mathbf{k}\in \mathrm{MBZ}}
(\mathbf{H}_{\mathbf{k}})_{nm}\langle nm\rangle_{\mathbf{k}},\end{align}
then derive the saddle-point equations of the mean-field parameters $n_f$, $m_f$, $h$, $\eta$ and chemical potential $\mu$ from knowledge of the elements of $\mathbf{H}_{\mathbf{k}}$. For example, the equation from $\partial F/\partial h=0$ is derived as
\begin{align}
m_f-\frac{1}{N}\sum_{\mathbf{k}\in \mathrm{MBZ}}(\langle 55\rangle_{\mathbf{k}}+\langle 88\rangle_{\mathbf{k}}-\langle 66\rangle_{\mathbf{k}}-\langle 77\rangle_{\mathbf{k}})=0,
\end{align}
the equation from $\partial F/\partial \eta=0$ is
\begin{align}
n_f-\frac{1}{N}\sum_{\mathbf{k}\in \mathrm{MBZ}}(\langle 55\rangle_{\mathbf{k}}+\langle 66\rangle_{\mathbf{k}}+\langle 77\rangle_{\mathbf{k}}+\langle 88\rangle_{\mathbf{k}})=0,
\end{align}
and equation from $\partial F/\partial \mu=-n_t$ is
\begin{align}
n_t-\frac{1}{N}\sum_{\mathbf{k}\in \mathrm{MBZ}}\sum^8_{i=1}\langle ii\rangle_{\mathbf{k}}=0.
\end{align}
The other two equations corresponding to $\partial F/\partial n_f=0$ and $\partial F/\partial m_f=0$ have much complex expressions. In this paper, we restrict the discussion to $n_t=2$.

\section{the saddle-point equations for PM phases}

For PM phase, the Hamiltonian matrix Eq. (\ref{VZ}) can be easily diagonalized analytically, leading to the expression for dispersions $E^\pm_{\mathbf{k}}$ and ground-state energy $E^{PM}_g$ in and above Eq. (\ref{Eplusminus}). Then the saddle-point equations can be derived by minimizing of $E^{PM}_g$ with respect to $n_f$, $\mu$ and $\eta$, to obtain
\begin{align} &n_t=\frac{2}{N}\sum_{\mathbf{k},\pm}\Theta(-E^\pm_{\mathbf{k}})\nonumber\\
&n_f=\frac{1}{N}\sum_{\mathbf{k},\pm}\Theta(-E^\pm_{\mathbf{k}})
[1\mp\frac{\epsilon^d_\mathbf{k}-\tilde{\epsilon}^f_\mathbf{k}}{\sqrt{(\epsilon^d_\mathbf{k}-\tilde{\epsilon}^f_\mathbf{k})^2+4V^2Z^2S^2_\mathbf{k}}}]\nonumber\\
&\eta=\frac{2Z}{N}\frac{\partial Z}{\partial n_f}\sum_{\mathbf{k},\pm}\Theta(-E^\pm_{\mathbf{k}})
[\epsilon^f_\mathbf{k}\pm\frac{2V^2S^2_\mathbf{k}-\epsilon^f_\mathbf{k}(\epsilon^d_\mathbf{k}-\tilde{\epsilon}^f_\mathbf{k})}{\sqrt{(\epsilon^d_\mathbf{k}-\tilde{\epsilon}^f_\mathbf{k})^2+4V^2Z^2S^2_\mathbf{k}}}],
\end{align}
in which
\begin{align}
\frac{\partial Z}{\partial n_f}=\sqrt{\frac{1-n_f}{2(2-n_f)^3}}-\frac{1}{\sqrt{2(2-n_f)(1-n_f)}}.
\end{align}
The self-consistent equations of AF and PM phases should be solved by numerical iteration.

\end{appendices}

\end{document}